\documentclass{article}

\usepackage{arxiv}

\usepackage[utf8]{inputenc} 
\usepackage[T1]{fontenc}    
\usepackage{hyperref}       
\usepackage{url}            
\usepackage{booktabs}       
\usepackage{amsfonts}       
\usepackage{nicefrac}       
\usepackage{microtype}      
\usepackage{graphicx}
\usepackage[authoryear,round]{natbib} 
\usepackage{doi}
\usepackage{tabularx}
\usepackage{array}

\title{Building the Ipseome:\\Large, Free, Open, Human Identity Data}
\author{ \href{https://orcid.org/0000-0002-4140-0268}{\includegraphics[scale=0.06]{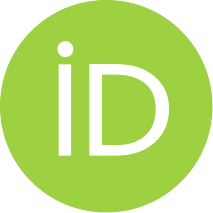}\hspace{1mm}Jason Jeffrey Jones}\thanks{\url{https://jasonjones.ninja} } \\
	Department of Sociology and Institute for Advanced Computational Science\\
	Stony Brook University\\
	Stony Brook, New York, USA \\
	\texttt{jason.j.jones@stonybrook.edu} \\
}

\renewcommand{\shorttitle}{Building the Ipseome}

\hypersetup{
pdftitle={Building the Ipseome},
pdfauthor={Jason Jeffrey Jones},
}

\begin{document}
\maketitle

\begin{abstract}
Shared data accelerates scientific progress.  Here, I describe the ipseome --- the largest free and open dataset on the topic of human identity.  The dataset is designed as reusable research infrastructure, with publicly accessible data repositories, documented measurement procedures, and versioned files for cumulative research on identity.  First, I present the motivation for and the ipseological principles driving construction of the ipseome.  Then, each component is introduced and discussed.  Finally, I summarize the current state of progress toward the ultimate goal.
\end{abstract}

Sporadically, individuals ponder a daunting question: \textit{Who am I?}.  Unfortunately for us, as scientists, they mostly do so lying in bed, unable to sleep at three in the morning; that makes observation of their responses inconvenient.  In this work, I present data I have collected in which individuals provide self-descriptions not lost in the haze of insomnia.

When individuals describe their selves with language, I call this \textit{personally expressed identity (PEI) text} \citep{jones_dataset_2021, jones_ipseology_2023}.  It is \textit{personal}: the authors are describing themselves.  It is \textit{expressed}: the authors’ words are available where others may see them.  It describes \textit{identity}: the explicit purpose of the text is description of the author.  The current project aims to document human selves as observed through personally expressed identity text.

Human selves are fascinating and complex.  Individuals vary over uncountable dimensions.  Each self is a unique pattern evolving over time.  They constantly invent, adopt, and reject expressions of identity.  For those who share my fascination with human selves, I am building the \textit{ipseome}---a public good for human identity research.

Despite their dynamism and variety, there is structure to human selves.  I know there is structure because I have seen it, documented it.  Some identities last longer than others \citep{vahabli_identity_2025}.  Components of identity - such as politics \citep{rogers_using_2021} and sexuality \citep{jones_lgbtq_2024} - change in salience over time.  Identities cluster in social networks \citep{maier_online_2026,tucker_pronoun_2023}.  Residents of different nations vary in their patterns of self-expression \citep{handzlik_hineni_2024}.  In this past work, I have introduced and developed \textit{ipseology} \citep{jones_ipseology_2023}.

Ipseology is the study of ipseity - defined as selfhood, individuality and the elements of identity.  I want to study ipseity at scale, so I have developed computational social science methods and compiled large datasets.  In this manuscript, I introduce the resources necessary for a larger community of ipseology.  \textbf{It is my hope that every scholar of the human self will find here a path to answers they seek.}

\section{Ipseome: What is it?  Why build it?}

\subsection{What is the ipseome?}

An -ome refers to a comprehensive catalog of the thing that precedes the suffix.  The goal of the Human Genome Project was to identify all human genes.  The proteome is the entire set of proteins produced by an organism.  A connectome is a comprehensive map of synaptic connections in an organism's nervous system.  The Latin \textit{ipse} has several meanings, all related to \textit{self}.  Thus, the ipseome refers to a comprehensive cataloging of self.

We have a rich and complex understanding of the constituent parts of human cells \citep{manzoni_genome_2018}.  In contrast, we have almost no data – and thus, a very poor model – of the elements that combine to form human selves.  Omes are painstakingly-compiled systematic descriptions.  I will construct the ipseome.  The goal is to create a large, consistently and persistently collected dataset of expressions of self.

How would you catalog an entire human self?  And then the population of selves?  I advocate an ipseological approach \citep{jones_ipseology_2023} based on these principles:

\begin{enumerate}
  \item Language is the data.
  \item Start with self-authored self-descriptions.
  \item Center time.
  \item More data is better.
  \item Quantitative description first; theory later.
\end{enumerate}

\subsubsection{Language is the data.}

Individuals construct their selves by telling stories.  They use language to introduce themselves.  To understand human selves, we should meet their terms---literally.  The best starting point for studying identity is to collect language that describes selves.

This has been the implicit assumption of many before.  Self-authored self-descriptions are easy to elicit and provide rich, deep objects for study.  \citet{kuhn_empirical_1954} introduced the Twenty Statements Test---a simple survey that asked individuals to compose twenty statements describing themselves.  Since then, many have adopted (e.g. \citep{griffo_who_2021}) and adapted (e.g. \citep{wylie_self_1961}) the TST.


\subsubsection{Start with self-authored self-descriptions.}

A self-authored self-description is a rich object for study.  The subject has chosen their own words, ordered, and structured them.  Moreover, they've chosen what to omit.  Salience and hierarchy \citep{stryker_symbolic_1980} are revealed by these choices.

Individuals include \textit{identity signifiers} within self-authored self-descriptions.  An identity signifier is a symbol chosen by an individual to represent an aspect of the self.  In text, signifiers are linguistic tokens - words, phrases, emojis, abbreviations, etc.  One may pre-define categories of signifier (e.g. political identities such as \texttt{conservatitve} and \texttt{liberal}) and seek them out.  Or, one may take an "open vocabulary" approach \citep{schwartz_personality_2013}.  In an open vocabulary approach, any and all signifiers are cataloged.

Self-authored self-descriptions exist in many forms.  Informally, individuals introduce themselves with a few words or sentences when they make a new acquaintance.  Job seekers write a dense, promotional self-summary at the top of their \textit{curricula vitae}.  Book authors shape how readers see them by composing an \textit{About the Author} passage for the dust jacket.

Most numerously available by far, however, are the bios that individuals write to be displayed on their social media profiles.  On Twitter alone, millions of individuals composed and revised publicly observable self-authored self-descriptions over the course of more than a decade \cite{handzlik_hineni_2024}.

The popularity of Twitter and other social media also provides another benefit: greater demographic coverage of populations than previously collected self-description data.  \citet{rhee_spontaneous_1995} is a highly-cited article comparing cross-national tendencies in self-descriptive text.  Without even an acknowledgement of the limited sample, however, a small set of New York University psychology-major undergraduates stood in for all Americans. By contrast, a substantial portion of American adults (twenty to twenty-five percent) report having a Twitter account \citep{pew_research_center_social_2024}.

\subsubsection{Center time.}

Individuals and the societies in which they are embedded are dynamic.  They change over time.  All claims about identity should address the temporal dimension.  We should know whether we have found something true or \textit{true today}.

In ipseology, we meet this mandate by centering time.  Data collection is never finished.  Every analysis assumes tomorrow's data will soon be added to today's.  We believe persistent, consistent measurement provides compounding value.

\subsubsection{More data is better.}

In ipseology, we seek data at scale.  This means many observations over long temporal periods and wide geographic dispersion.  Precise estimates are necessary to make sound decisions in the face of uncertainty.  The fact that more observations lead to more precise estimates is frequently taught in introductory statistics and has even been tested and verified empirically \citep{asiamah_larger_2017}.  Longitudinal and cross-sectional sampling over time reveals both individual change and cultural evolution.  No matter where one lives, one can answer the question "Who am I?"  The more answers gathered from around the globe, the better for future understanding of the human self.

\subsubsection{Quantitative description first; theory later.}

Ipseology starts from the perspective that previous theories of identity have been verbal, intuitive narratives born in a data-starved era.  They may turn out to be true or useful, but we won’t know until after comparison to good data.  The current social science paradigm demands theory as a starting point and data for confirmation.  In ipseology, we reverse the order.  Data first; we can circle back to theories and model-making after they can be constrained by consistent, persistent observation.

The principles above animate all of the efforts I describe below.  Continuing with these priniciples will lead us closer to the ipseome.

\subsection{Why build the ipseome?}


As scholars of identity, we have the luxury of studying something everyone is interested in: themselves.  Science and society are starved for information about who we are, why, and how we change.  The ipseome is a first and significant step toward answers.  More broadly, this project will serve as a model for modern social science.  Consistent, persistent data collection at scale is the rocket that social science needs to launch past the replication crisis and into a modern space of sound practice.

More concretely, the compelling reason to build the ipseome is this: \textbf{shared data advances fields of study}.

Free and open datasets are widely available for academic studies of computer vision and natural language processing.  I believe it is not a coincidence these fields of study have made rapid, recent progress.  The availability of these datasets has been described as "central," "essential," and "crucial" by researchers \citep{emam_state_2021,lhoest_datasets_2021}.  I aim to bring open data to the study of personally expressed identity text.  

Among other benefits, free and open data lowers the barrier to entry.  The resources described below are too large and complex for one researcher or lab to monopolize.  I offer them as free public goods.  Any researcher can download existing data and apply it to their own research questions.

\section{Ipseity Daily}

Ipseity Daily\footnote{\url{https://jasonjones.ninja/social-science-dashboard-inator/ipseity-daily/}} is a system to continuously collect and distribute human identity observations.  On a daily cadence, an online survey is fielded to paid, randomly-selected respondents.  Every item on the survey takes this form: \textit{Does \texttt{<signifier>} describe you today?}  In place of \texttt{<signifier>} is a word, phrase or emoji - e.g. \texttt{dad}, \texttt{Scorpio}, \texttt{Cleveland Browns fan}, \texttt{science nerd}, \texttt{:crown:}, \texttt{:smiling-face-with-hearts:}.  Each respondent chooses Yes, No or Skip for many randomly selected signifiers.

At the time of writing, 707 unique signifiers are eligible and 80 are chosen for each respondent.  Each day, 21 respondents are recruited.  These numbers may change.

The sponsor of Ipseity Daily is \href{https://jasonjones.ninja/jason-jeffrey-jones-productions/}{Jason Jeffrey Jones Productions}, and the organization is committed to fund at least one million observations.  At the current rate of 21 respondents x 80 observations per respondent, Ipseity Daily is expected to continue through February 24th, 2027.

\paragraph{Data availability.}
All data from Ipseity Daily is available from the \href{https://jasonjones.ninja/social-science-dashboard-inator/ipseity-daily/download.html}{primary download page} or through a \href{https://zenodo.org/records/21126659}{Zenodo archival mirror}.

Data in this form addresses many research questions.  Here, I list several:

\begin{itemize}
  \item Which identities are most prevalent today?
  \item At what rate is prevalence changing for each identity signifier?
  \item Can any person, theory or system predict the prevalence of signifiers one month in the future?  One year?
  \item Which sets of signifiers cluster in individuals and which sets preclude others?
  \item Which signifiers endure and which are ephemeral?  (By chance, respondents will be resampled on different dates, allowing some degree of longitudinal analysis.)
\end{itemize}

It is my firm belief that \textit{any} research question in the realm of human identity could be addressed through the \href{https://jasonjones.ninja/social-science-dashboard-inator/ipseity-daily/}{Ipseity Daily} process.  Because it is mostly software, the process can scale with funding to novel targets of study - i.e. new signifiers - and greater precision - i.e. more daily respondents.  More broadly, the process can be copied in other nations to eliminate the current limitation of US-only coverage.




\section{JJJ Pro Who am I?}

\textit{JJJ Pro Who am I?} is a multiwave survey that elicits self-authored self-descriptions from a representative sample of US adults.  In each wave, half of the respondents are re-recruited from the previous sample to allow year-over-year longitudinal analysis at the individual level.  The other half of the respondents comprise a fresh representative sample to allow continuous cross-sectional analysis at the population level.  I believe the Jason Jeffrey Jones Productions Who am I Dataset (\textit{JJJ Pro Who am I?}) is the \textbf{first and only free, open, representative-sample, full-text self-description data ever made available}. 

To construct the dataset, I used a prompt similar to the widely adopted Twenty Statements Test (TST; \citep{kuhn_empirical_1954}), and I informed respondents how their responses would be shared.  See the section Methodology Details for a comprehensive account of the respondents' experience.

\subsection{The Data}

For convenience, the data is a single file containing one row per observation.  Each row contains a self-authored self-description, the date of observation, a one-way-hashed anonymized respondent ID, and the available demographics for the respondent.  The file is available for download at no cost (free) through a Creative Commons Attribution, Non-Commercial, Share-Alike license (Creative Commons, 2024).

Respondents were informed during consent and before responding that the data would be made publicly available.  See the section Methodology Details - Procedure for details.

\paragraph{Data availability.}
All data from JJJ Pro Who am I? is available from the \href{https://osf.io/74asx/files/5ybuv}{primary download page} or through a \href{https://doi.org/10.5281/zenodo.21134632}{Zenodo archival mirror}.

\subsection{Methodology Details}

\subsubsection{Procedure}

In this section, I will guide you through the procedure chronologically as respondents experienced the survey.

\begin{enumerate}
  \item Through the Prolific platform \citep{eyal_data_2021,palan_prolific_2018,peer_beyond_2017} respondents were offered the opportunity to participate in a survey with the title "A Survey of Americans 2024b" and the description "In this study, you will be asked to write about yourself."
  \item After clicking a button to open the study, the respondent saw a screen requesting consent.  The full text is available in this document in the section Consent Screen.
  \item If the respondent chose "Do NOT Consent," the survey ended and the following text was displayed: "As you do not wish to participate in this study, please close this survey and return your submission on Prolific by selecting the 'Stop without completing' button."
  \item After choosing "Consent to continue," the respondent saw the PEI definition screen.  See the full text in the section PEI Definition Screen.  The survey would not continue until and unless the respondent chose the correct response: "Personally expressed identity text refers to self-authored self-descriptions."
  \item After correctly responding, the respondent saw the data sharing information screen.  See the full text in the section Data Sharing Screen.  The survey would not continue until and unless the respondent chose the correct response: "Your data will be available for public download under a Creative Commons license."
  \item After correctly responding, the respondent saw the \texttt{wai\_text\_response} item.  See the full text in the section Who am I? Screen.  Respondents typed their responses in a multiline text area.
  \item Upon submitting a response to the \texttt{wai\_text\_response} item, the respondent was redirected back to the Prolific platform with a completion code.
\end{enumerate}

\subsubsection{Payment}

Each respondent was paid \$1.00.  The total cost of the study was \$611 dollars.  (I requested 610 responses and was delivered 611.)  Jason Jeffrey Jones Productions paid the entire cost.  Jason Jeffrey Jones Productions is a negative-profit entity that converts portions of my salary to free public goods to accelerate research.

The median time to completion was six minutes and twenty-eight seconds.  The rate for participant payment was thus about \$9.28 per hour.

\subsubsection{Sample Demographics}

To demonstrate demographic coverage, here I include a cross-tabulation of Age and Sex and a simple frequency table for Ethnicity for both extant waves.  Recall that all microdata is available for download.





\subsubsection{Names Removed}

In exactly three cases, the respondent included a sentence providing their name within their response to the \texttt{wai\_text\_response} item.  (This was in contradiction to the second bullet of the prompt: Do NOT use any name or number that is unique to you.  In one case the respondent provided initials, and in the other cases only a first name, so it is arguable they felt they were complying.)  I chose to remove the sentence that indicated a name from all three responses.  This is the only editing I made to the responses.  No responses were excluded.

\subsubsection{Responses Read}

I read the entirety of every response.  I did so to ensure conformance with my intention to release data that did not uniquely identify any individual.  It is my belief (after removing names, as described above) that no response uniquely identifies anyone.

I have chosen not to remove any "sensitive" responses.  Some responses discuss trauma, including abuse.  That was the respondent's choice, after being well-informed about the purpose of the study.  It is my opinion that withholding data to avoid offending sensibilities would cause more harm than good.  Any researcher is welcome to filter and re-release the data according to their own criteria.

I have chosen not to remove any "invalid" responses.  Two responses are quite obviously the output of a large language model chatbot.  Other responses are suspiciously unlike the rest of the responses and potentially chatbot-generated.  As of this moment in 2024, there is no reliable method to separate machine from human generated text.  It is my belief there never will be.  Removing chatbot responses would introduce the problems of false-negative and false-positive removals.  Including all responses (as I have chosen) provides researchers a chance to estimate the frequency of automated or other "invalid" responses by their own criteria.

Many respondents did not provide exactly twenty responses.  Again, I chose to include all responses - whether a two-word answer, a numbered list of twenty words or a multi-paragraph stream-of-consciousness composition.  There is information to be gleaned from the format, length and content of responses that did not strictly comply with the prompt.







\section{HINENI}

For over a decade, hundreds of millions of individuals around the world described themselves within their Twitter profile biographies.  From 2012 through 2023, I used the Twitter Streaming API to collect self-authored self-descriptions at scale.  Specifically, I observed the profile biographies for the authors of a one-percent random sample of all tweets.

Human Identities across Nations of the Earth Ngram Investigator (HINENI) was built from these observations \citep{handzlik_hineni_2024}.  Individual accounts were mapped to nations depending on their stated locations (e.g. Rio de Janeiro → Brazil, Columbus, OH → USA).  An annual census was built by collecting all accounts within a nation and randomly sampling one bio observation per individual each year.  The data contains the incidence and prevalence of identity signifiers per nation and year.


HINENI is most easily explored through the dashboard at \url{https://jasonjones.ninja/hineni/}.  This interactive web application has a familiar search-like interface.  It is easy to query identity signifier prevalence over multiple nations and years.

\paragraph{Data availability.}
All data from HINENI is available from the \href{https://jasonjones.ninja/hineni/data.html}{primary download page} or through a \href{https://osf.io/download/k7bwj/}{Open Science Foundation archival mirror}.

HINENI provides immediate access to the relative popularity of self-ascribed identities for thirty-two nations over twelve years.  \textbf{There is no public human identity data that offers broader temporal and geographic coverage.}

Previously, my work has shown that \textit{Americans} increasingly politicized their self-concepts.  More and more often they chose political words (e.g. liberal and conservative) to describe themselves \citep{rogers_using_2021}.  HINENI data demonstrates this was a global phenomenon; over the last decade, individual self-descriptions became increasingly politicized in 30 out of 32 nations with adequate data \citep{clemente_online_2024}.


\section{Jason Jeffrey Jones Identity Trends V2}

\textit{Jason Jeffrey Jones Identity Trends V2} provides annual estimates of identity signifier prevalence in the United States.  It is a predecessor to HINENI and is built on Twitter profile biographies of only those users who displayed a US place name as their location.

\citet{jones_dataset_2021} documents the methods used to collect, filter and parse US Twitter profiles.  The same methods were used to create this updated version (V2), which includes data through 2023.

The online dashboard at \url{https://jasonjones.ninja/jason-jeffrey-jones-identity-trends-v2/} offers a no-download, no-code starting point for exploration.

\paragraph{Data availability.}
All data from Jason Jeffrey Jones Identity Trends V2 is available from the \href{https://jasonjones.ninja/jason-jeffrey-jones-identity-trends-v2/about.html}{primary download page} or through a \href{https://osf.io/ed3kj/files/z7b8j}{Open Science Foundation archival mirror}.

\section{Words You Today}

Words You Today\footnote{\url{https://jasonjones.ninja/words-you-today/}} is a Web application individuals use to track and compare their identity signifiers over time.  Users see one signifier at a time and respond to the question \textit{Does this describe you today?} with Yes, No or Skip.

An anonymized dataset will be freely available for research.

\section{Discussion}

The ipseome records individual selves at scale.  The ipseome centers the fact that humans change over time. It acknowledges that observations are merely low-fidelity snapshots of a multifaceted subject in constant motion.  The five principles of the ipseological approach necessarily follow.

I have taken several approaches to data collection, and therefore the ipseome is a collection of datasets.  Table~\ref{tab:ipseome-components} lists and describes each component.

\begin{table}[htbp]
\centering
\caption{Components of the Ipseome}
\label{tab:ipseome-components}
\small
\begin{tabularx}{\textwidth}{
  >{\raggedright\arraybackslash}p{0.22\textwidth}
  >{\raggedright\arraybackslash}X
  >{\raggedright\arraybackslash}p{0.18\textwidth}
  >{\raggedright\arraybackslash}p{0.16\textwidth}
}
\toprule
Ipseome Component & Goal & Website & Citation \\
\midrule

\textit{Ipseity Daily}
&
Consistently and persistently measure the prevalence of identity signifiers with a repeated, cross-sectional survey in American adult samples.
&
\href{https://jasonjones.ninja/social-science-dashboard-inator/ipseity-daily/}{Website}
&
In preparation
\\

\textit{JJJ Pro Who am I?}
&
Consistently and persistently compile responses to the Twenty Statements Test. In each annual wave, both a new cross-sectional representative sample and a longitudinal repeated-subjects sample are collected.
&
In preparation
&
In preparation
\\

\textit{HINENI}
&
Estimate identity signifier prevalence at annual cadence within cross-sectional national samples of active Twitter users.
&
\href{https://jasonjones.ninja/hineni/}{Website}
&
\cite{handzlik_hineni_2024}
\\

\textit{JJJITV2}
&
Estimate identity signifier prevalence at annual cadence within a sample of active United States Twitter users.
&
\href{https://jasonjones.ninja/jason-jeffrey-jones-identity-trends-v2/}{Website}
&
\cite{jones_dataset_2021}
\\

\textit{Words You Today}
&
Bring ipseology into individuals' lives in a user-directed and constructive manner.
&
\href{https://jasonjones.ninja/words-you-today/}{Website}
&
In preparation
\\

\bottomrule
\end{tabularx}
\end{table}

\subsection{Conclusion}

With an extant and growing ipseome, a new type of systematic science becomes possible.  Suddenly, the study of human selves transforms into a science similar to meteorology or astronomy.  Theories must bend to the facts revealed by daily observation.  Models of the human self and its development over time can finally be tested against data.

\bibliographystyle{unsrtnat}
\bibliography{Ipseome}

\end{document}